\documentstyle[]{l-aa}
\input psfig.tex
\def\lsim{\lower.5ex\hbox{$\; \buildrel < \over \sim \;$}}
\def\gsim{\lower.5ex\hbox{$\; \buildrel > \over \sim \;$}}
\def \simeq{\lower.3ex\hbox{$\; \buildrel \sim \over - \;$}}
\def\ch{\lower-0.55ex\hbox{--}\kern-0.55em{\lower0.15ex\hbox{$h$}}}
\def\lh{\lower-0.55ex\hbox{--}\kern-0.55em{\lower0.15ex\hbox{$\lambda$}}}

.tfm

\begin{document}
\title{Class Transitions and Two Component Accretion Flow in GRS 1915+105}
\author{Sandip K. Chakrabarti$^{1,2}$, Anuj Nandi$^{1}$, A. K. Chatterjee$^{3}$, A. K. Choudhury$^{3}$ and U. Chatterjee$^{3}$}
\institute{$^1$ S. N. Bose National Centre for Basic Sciences, Salt Lake, Kolkata 700098, India\\
$^2$ Centre for Space Physics, Chalantika 43, Garia Station Rd., Kolkata, 700084, India\\
$^3$ Centre for Space Physics (Malda Branch), Atul Market, Malda 732101, India\\
e-mails:chakraba@bose.res.in, space\_phys@vsnl.com}
\offprints {S. K. Chakrabarti {\it chakraba@boson.bose.res.in}}
\date{Received ; accepted , }
\maketitle
\markboth{  }{}

\begin{abstract}

The light curve of the galactic micro-quasar GRS 1915+105 changes in at least 
thirteen different ways which are called classes. We present examples of the 
transitions from one class to another as observed by the IXAE instrument 
aboard the Indian Satellite IRS-P3. We find that the transitions are associated 
with changes in photon counts over a time-scale of only a few hours and  they
take place through unknown classes. Assuming that the transitions are caused by
variation of the accretion rates, this implies that a significant fraction of the matter must be 
nearly freely falling in order to have such dramatic changes in such a short time.
\end{abstract}

\noindent ACCEPTED FOR PUBLICATION IN ASTRONOMY AND ASTROPHYSICS

\section{Introduction}

GRS 1915+105 is well known for its diversity of light curves (e.g. Morgan, 
Remillard \& Greiner, 1997; Belloni et al. 2000). RXTE has pointed at 
it numerous times and yet the light curve has remained 
largely unpredictable. Belloni et al. (2000), in a model independent 
way, classified most of the light curves into twelve classes 
which are designated as $\chi,~ \alpha,~ \nu, ~ \beta, ~ \lambda, 
~ \kappa, ~ \rho, ~ \mu, ~ \theta , ~ \delta , ~ \gamma$ and $\phi$.
Naik et al. (2002a) showed that there is another independent class called
$\omega$. Alhough the light curve was observed to change from one class to another, 
the actual transition was never reported and therefore, the actual physical
process which triggers a specific class transition has never been investigated.

It was predicted in several earlier papers using the advective flow paradigm
(Chakrabarti \& Nandi, 2000; Nandi, Manickam \& Chakrabarti, 2000; Chakrabarti 
et al. 2002) that variation of the Keplerian and the sub-Keplerian accretion
rates might cause class transitions. It was pointed out that 
there are actually five fundamental states differing only by Keplerian 
and sub-Keplerian accretion rates. Ways in which the transition occurs 
between these states decide which class would be seen.  It was also pointed 
out that the outflows play a major role in class transitions, since they interact 
with the soft photons and affect the spectral slopes as well.
Recently, Chakrabarti et al.  (2004) presented two examples of class transitions 
from IXAE observations and concluded that a class transition always 
take place through some unknown class. In many of the classes that GRS 1915+105 
exhibits, one could see the presence of Quasi-Periodic Oscillations (QPOs).
Very recently, using extensive time-dependent numerical simulation of accretion
flows that include cooling effects, it has been shown (Chakrabarti, Acharyya \& 
Molteni, 2004) that the so-called advective disk paradigm is capable of explaining 
QPOs very naturally. The prime cause of the quasi periodic oscillations (QPOs) 
of X-rays from compact objects is found to be quasi-coherent shock 
oscillations. The post-shock region (i.e., the so-called CENtrifugal pressure 
supported BOundary Layer, or CENBOL) outside a black hole horizon acts 
as the Compton cloud by intercepting soft photons from a Keplerian disk and 
reprocessing them to high energies.  Along with the shock oscillations, 
the size of the CENBOL changes and therefore the number of intercepted soft photons 
oscillates, causing the observed QPO. Power density spectra of these `simulated' light curves
directly show how QPOs occur at or near break frequency -- a well-known observed 
phenomenon. The frequency of oscillation is thought to be related to the inverse 
of the infall time-scale (Molteni, Sponholz and Chakrabarti, 1996) and as 
such should increase with the increase of the sub-Keplerian accretion rate 
undergoing the shock transition as the cooling rate is increased. This general behaviour
has also been observed (Remillard et al. 1999).

In this paper, we present a large number of examples of the `rare' class transitions, 
all of them being from the Indian satellite data, and analyze what happens during 
such a transition. In particular, we follow the light-curve, the power density spectra 
and the photon spectra throughout the transition. We found that: (a) A class 
transition is invariably accompanied by a significant variation of the average X-ray photon 
count rate, indicating that either the Keplerian disk rate, or the sub-Keplerian 
flow rate or both may be changing, (b) In between two known classes, a class of 
unknown type appears for hundreds of minutes and (c) During a transition, 
the photon index becomes noisy until the flows settles into 
a new class indicating the presence of turbulent behaviour during transition.
At the same time, we also study the behaviour of QPOs and show how the frequency 
is changed consistently with the accretion rates as inferred from the spectra.
In the next Section, we present the observational results on class 
transition. Based on the new inputs from the observational
results, in Section 3, we discuss what the nature of the accretion flows might be.
We find that in order to enable class transition in a few hours,
a significant fraction  of the flow must be nearly freely falling, i.e., sub-Keplerian.
Finally, in Section 4, we draw our conclusions. 

\section{Observation of Class Transitions}

The results we discuss in this section were obtained by the Pointed Proportional Counters (PPCs)
in the IXAE instrument aboard the Indian Satellite IRS-P3 (Agrawal 1998) which functioned 
during 1996-2000. The operating energy range is between 2 and 18 keV. The counts are
saved in the archive only in two channels -- one is 2-6keV and the other is 6-18keV. 
The time resolution in Medium mode could be $0.1$s but normally the time resolution was
set to be $1$s. As a result of the presence of only two energy 
channels, only two points could be obtained in the spectrum and 
a so-called `mean photon index' (MPI) $s_\phi$ can be calculated after each second. 
Similarly, $0.1$s time resolution in the Medium mode restricts the 
observation of QPOs up to $5$Hz only, while the $1$s resolution restricts the 
reporting of QPOs up to $0.5$Hz only. Nevertheless, the light curves are clear
enough and the identification of the specific class can be done without ambiguity
(e.g., Naik et al 2001, 2002ab; Paul et al. 2001). It is to be noted that
(a) the counts in the high energy bin could be very low and the $s_\phi$ suffers 
from low number statistics and (b) the slope itself is known to vary
in the $2-18$keV range, especially, the spectrum becomes harder above $10-12$keV in the hard state. 
An assumption of a constant slope $s_\phi$ will thus be
erroneous. Therefore, while $s_\phi$ gives an indication of how the slope changes (as a colour-colour
diagram) its absolute value should be treated with caution.

Before we present the IXAE observations, it is useful to 
give a brief description of the QPOs which are so far observed in GRS 1915+105.
Broadly speaking, this can be subdivided into four classes:
(i) low frequency QPO (LFQPO) in the range $\sim 0.001- 0.02$ Hz, (ii) break
frequency (BF) or intermediate frequency QPO (IFQPO) in the range $\sim 0.1-0.3$ Hz,
(iii) high frequency QPO (HFQPO) in the range $\sim 1-10$ Hz and (iv) the
very high frequency QPO (VHFQPO) around 67 Hz.

\begin{figure}
\vbox{
\vskip -3.0cm
\hskip 0.0cm
\centerline{
\psfig{figure=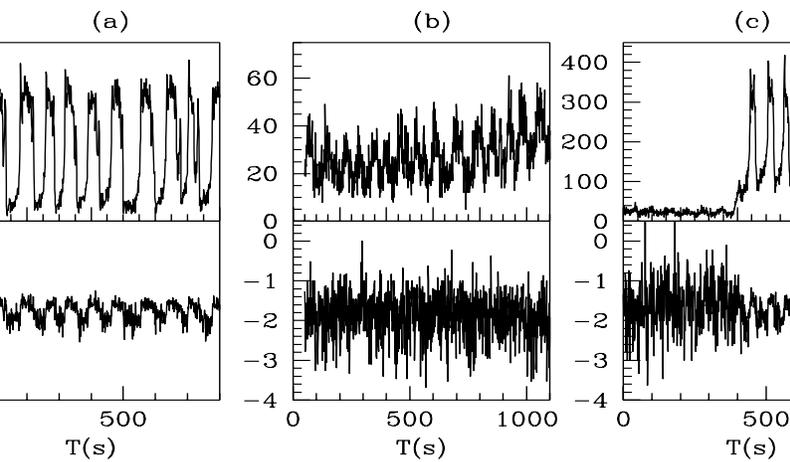,height=13truecm,width=16truecm}}}
\vspace{-2.0cm}
\caption[] {$2-18$keV light curves as observed by IXAE (upper panel)
and the mean photon spectral index $s_\phi$ (lower panel) in 1st, 3rd and 5th orbits
of June 22nd, 1997 (see, Table 1). GRS1915+105 was (a) in the $\kappa$ class, (b) in an unknown class
and (c) went to the $\rho$ class on that day. Lower panels show how $s_\phi$ distinctly 
change. Specifically it is noisy during the transition.
}
\end{figure}

In Table 1, we present the log of the observations we report in this paper which 
showed class transitions. The first column refers to the figure where the 
results are shown. The second column refers to the name of the Satellite. The 
third column shows the date of observation and the time when the observation started.
The fourth column gives the orbit numbers (or, the Obs. ID in case of RXTE)
plotted in the Figure. Typically, the time interval
between two successive orbits is around $80$ minutes. The fifth column gives the 
exact nature of class transition. Since during transition, a given  class
is not found to be `canonical' as defined by Belloni et al. (2000), we have put
the class-names inside quotation marks. 

\begin{table}
\small
\centering
\caption{\label{table1} Class transitions of GRS 1915+105 reported in this paper}
\vskip 0.5cm
\begin{tabular}{|c|c|c|c|c|}
\hline
FIGURE & Satellite & Date & Orbit No. & Class transition   \\ 
 &    & Time (UT) & ObsID &     \\ \hline

1. & IXAE & 22th June, 1997 & 1, 3, 5  & $\kappa$ $\rightarrow$ $\rho$ \\  \cline{3-3} 
  &  & 12:12$^a$ &  &   \\ \hline

2. & IXAE & 25th June, 1997 & 3, 4, 5  & `$\chi$' $\rightarrow$ $\rho$ \\  \cline{3-3} 
  &  & 11:12$^a$ &  &   \\ \hline

3. & IXAE & 08th June, 1999 & 2, 3  & `$\chi$' $\rightarrow$ $\theta$ \\  \cline{3-3} 
  &   & 11:02$^a$ &  &   \\ \hline

3. & RXTE & 08th June, 1999 & 40702-01-03-00  & `$\chi$' $\rightarrow$ $\theta$ \\  \cline{3-3} 
  &   & 13:52 &  &   \\ \hline

5. & IXAE & 25th June, 2000 & 2, 3  & `$\rho$' $\rightarrow$ $\alpha$ \\  \cline{3-3} 
  &  & 14:07$^b$ &  &   \\ \hline
\end{tabular}

\noindent {\small $a)$ Observation time at the begining of the first orbit;
$b)$ Observation time at the begining of the second orbit}\\
\end{table}

In Fig. 1(a-c), we present the light curves ($2-18$keV) of the June 22nd, 1997 observation in the upper panel
and the mean photon index (MPI) in the lower panels. The MPI $s_\phi$ is obtained using the definition:
$$
s_\phi=- \frac{log(N_{6-18}/E_2) - log(N_{2-6}/E_1)}{log(E_2)-log(E_1)} ,
\eqno{(1)}
$$
where $N_{2-6}$ and $N_{6-18}$ are the photon count rate from the top layer of the PPC
and $E_1$ and $E_2$ are the mean energies in each channel.
Thus, $E_1=4$keV and $E_2=12$keV respectively. We have thus normalized the count rate
per keV and then obtained the slope in the log-log plot since we expect a 
power-law slope in the $4-12$ keV range. The panel 1a is in the so-called $k$ class 
(Belloni et al., 2000). The panel 1b is in a unknown class and the panel 
1c clearly shows the transition from the unknown class to the so-called $\rho$ class. The panels
are separated by about three hours.  

In the lower panels, the $s_\phi$ oscillates between $\sim 2.4$ 
to $\sim 1.4$ in Fig. 1a very systematically. In Figs. 1b and 1c, the unknown class 
produced very noisy photon spectral slope variation. As soon as the $\rho$ 
class is achieved after one `semi-$\rho$' oscillation, noise in $s_\phi$ is 
reduced dramatically.

The IXAE observation of the 23rd, 24th and 25th of June, 1997 showed that the
system was still in $\rho$ class after the transition on 22nd of June, 1997. Subsequently,
on 25th June, 1997 there was another transition to $\chi$ and it returned back to $\rho$. 
It remained in $\rho$ class on the 26th of June before returning to $\kappa$ on the 
27th. Thus $\kappa \rightarrow <unknown> \rightarrow  \rho \rightarrow <unknown> \rho \rightarrow <unknown> \rightarrow
\kappa$ transitions took place in a matter of five days. The exact time and duration of the 
last transition mentioned above could not be seen because of the lack of observation. 

\begin{figure}
\vbox{
\vskip -3.0cm
\hskip 0.0cm
\centerline{
\psfig{figure=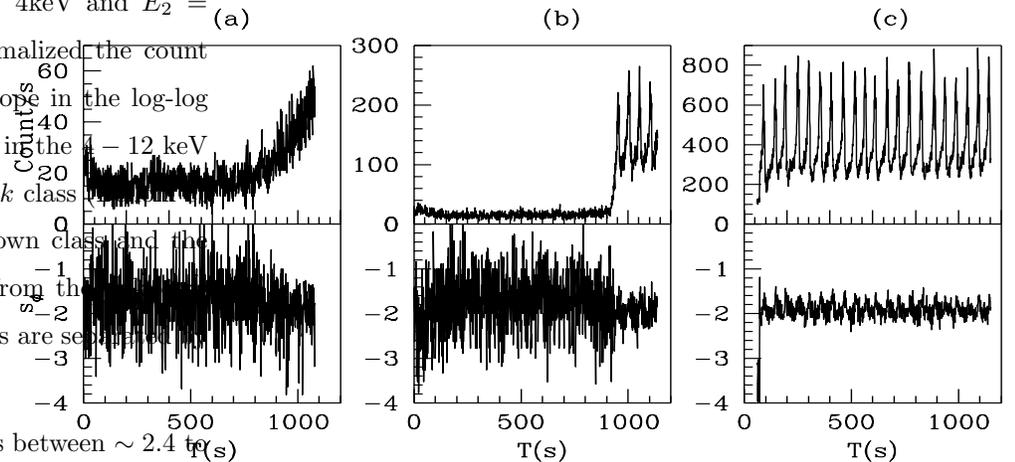,height=13truecm,width=16truecm}}}
\vspace{-2.0cm}
\caption[] { 
Class transition of GRS 1915+105 as observed by IXAE on the 25th of June, 1997 in three
successive orbits (See, Table 1). It was (a) in an unknown class similar to $\chi$, (b) in transition to  
$\rho$-type class with very low count rate and (c) in $\rho$ class after stabilization in that class.
In the lower panels are $s_\phi$ showing noisy behaviour during transition before settling
down in (c).
}
\end{figure}

In Fig. 2(a-c) the observation of IXAE on 25th of June, 1997 is presented. The panels are separated
by about one and a half hours. Here too, the upper and lower panels represent variation of 
photon count rates and  $s_\phi$ respectively. In Fig. 2a, the GRS 1915+105  is in the so-called 
$\chi$-like class, though the photon count rate showed considerable variations, not characteristic 
of $\chi$. Correspondingly, the average spectra also softened as is suggested by the gradual 
decrease in $s_\phi$. In Fig. 2b, this trend continued until a `semi-$\rho$' class 
was achieved and the noise in the photon spectra went down. In Fig. 2c, after one full 
orbit, the count rate went up further by about a factor of four and a steady  
$\rho$ state was achieved. The average (photon) spectral index $s_\phi$ was $\sim 1.75$ 
in Fig. 2a, but it became $\sim 1.9$ in Fig. 1c, indicating general softening. 

\begin{figure}
\vbox{
\vskip -5.0cm
\hskip 2.0cm
\centerline{
\psfig{figure=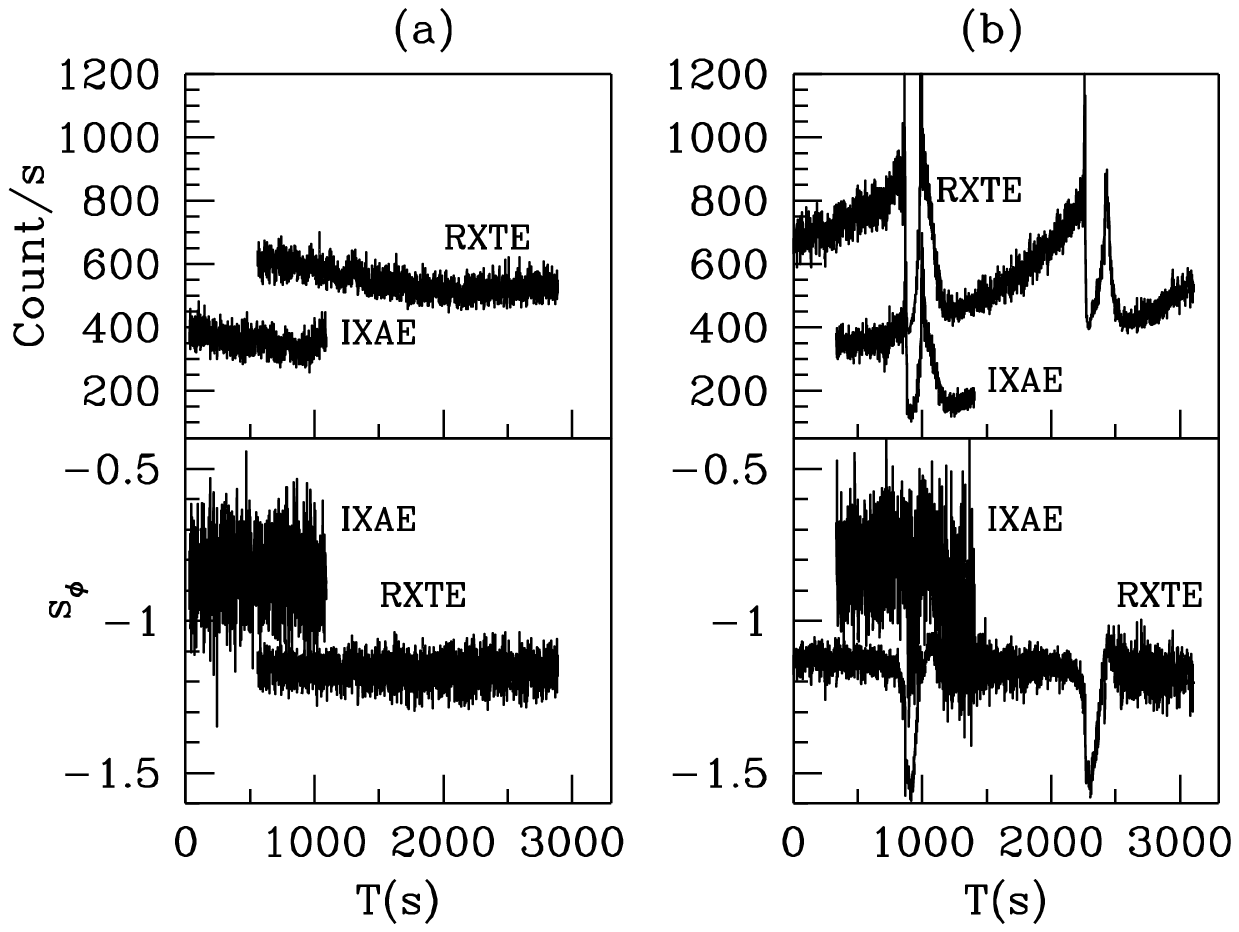,height=16truecm,width=16truecm}}}
\vspace{-2.5cm}
\caption[] {Class transition as seen from IXAE and 
RXTE observations on the 8th of June, 1999 in two successive orbits (marked). RXTE photon count rates are divided
by $50$ and shifted upward by $200/s$ for comparison. In (a), for a period of about $3000$s, there
was no significant variation in light curve or spectral index. The object was in a class similar to $\chi$
but the count rates were a factor of $10-20$ times higher. There is a gap of $44$ minutes in the two RXTE 
data presented in (a) and (b). In (b), the object is distinctly in the $\theta$ class. RXTE data is less noisy than the
IXAE data because of higher counts. It was binned in $2-6$ and $6-15$keV before computing $s_\phi$ so that
comparison with IXAE could be made.}
\end{figure}

In Fig. 3(a-b) we show the light curve and $s_\phi$ from IXAE data obtained on the 8th of June, 1999.
The two panels are from two successive orbits $\sim 80$ minutes apart. In Fig. 3a, the count rate
was very high compared to what is expected from a typical $\chi$ state although the 
power density spectrum (PDS) is typical of that of the $\chi$ class. A QPO at $4.7$Hz is present. 
The $s_\phi$ is $0.85$  which is harder than what is observed in Fig. 2. When combined with RXTE data of that
date (Fig. 3a), one finds that for a long time ($\sim 3000$s) there was no signature of any `dip' which is 
the characteristic of the $\theta$ class. Hence, this must be in an unknown class, more close to $\chi$ 
than any other. RXTE also observed this object on the 7th of June, 1999 and found the 
object to be in the $\chi$ class. In Fig. 3b, the light curve in the next orbit of IXAE
shows evidence of the so-called $\theta$ class. Interestingly, the spectra gradually `hardened'  
to $s_\phi \sim 0.6$ just before the `dip'. The spectra characteristically softened 
in the `dip' region with $s_\phi \sim 1.4$ as the inner edge of the disk 
disappeared. This class transition is confirmed in the data of RXTE also shown 
in Fig. 3b. The lower panels showed that the spectral slopes obtained for RXTE 
data calculated in a similar way to $s_\phi$ was calculated (Eqn. 1).
Here, the photons were first binned in $2-6$ keV and $6-15$keV (In epoch 4 of RXTE, the
science data is available in a maximum of 35 channels. Thus the energy channel width  could not be
made identical to $6-18$keV as in IXAE) before computing $s_\phi$ from
$$
s_\phi= - \frac{log(N_{6-15}/E_2) - log(N_{2-6}/E_1)}{log(E_3)-log(E_1)},
\eqno{(2)}
$$
where $E_1=4$keV, $E_2=9$keV and $E_3=10.5$keV. Note that there is a large difference between the
mean spectral slopes calculated from IXAE and RXTE data. The main reasons
appear to be (a) by decreasing $E_2$ from $18$keV in IXAE to $15$keV the 
value of $s_\phi$ is increased by $\sim 15$ per cent. 
(b) usually the spectrum in the low state (as in between the `dips' in the $\theta$ class)
has a harder tail for energy above $12$keV. Thus decresing the binsize limit from 
$18$keV to $15$keV decreases the photon counts in harder parts of the spectrum.
These two combined effects cause the mean spectrum of RXTE
to be softer. Its lower noise is clearly due to its very high count rates
(about $50$ times higher than IXAE) for its higher effective surface area
($6500 cm^2$ as opposed to $400 cm^2$ for one of the detectors of IXAE) 
and its usage of xenon as opposed to a mixture of $90$ per cent argon and $10$
per cent methane.

To show that the class in Fig. 3b is indeed that of 
the $\theta$ class, we plot in Fig. 4 the power density spectra. This showed a characteristic 
break at BF $\sim 0.1$Hz and the HFQPO at $5.4$Hz with a broad `Q' weak QPO at 
the break frequency. The PDS of the light curve (Fig. 3a) in the previous orbit 
is $\chi$-like and it does not show any LFQPO, BF/IFQPO or VHFQPO.

\begin{figure}
\vbox{
\vskip -3.0cm
\centerline{
\hskip 3.0cm
\psfig{figure=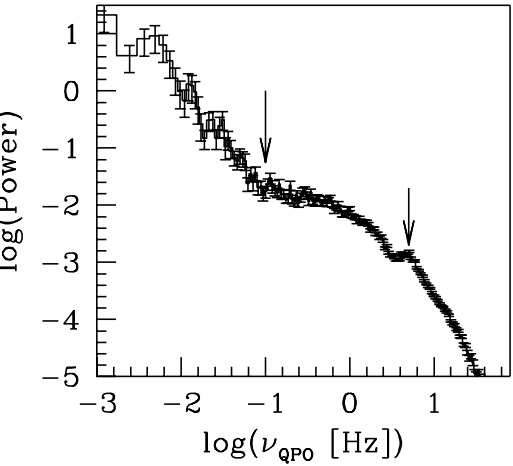,height=18truecm,width=18truecm}}}
\vspace{-5.0cm}
\caption[] {The power density spectrum of the RXTE light curve in (b). The
signature of $0.1$Hz break and a kink (weak and broad QPO) at around $5.4$Hz
are typical of a $\theta$ class. These, along with softening of the 
spectrum at the dips (prominent in both the data) indicating that a class transition 
has indeed taken place. }
\end{figure}

In Fig. 5(a-b), we show another example of a class transition in which the light curve in
the `$\rho$' class (Fig. 5a) goes over to the so-called $\alpha$ class (Fig. 5b). This is from the IXAE observation 
on the 25th of June, 2000. The count rate in this `$\rho$' class was much higher than that seen 
in Figs. 1 and 2 and the photon spectral index in the lower panel also showed that the spectra are harder 
(average $s_\phi \sim 0.65$ as compared to $\sim 1.8$ in Fig. 1 and $\sim 1.9$ in Fig. 2.). So, it could be 
an intermediate class. In the alpha class $s_\phi$ became noisy and the transition is clearly obvious. 
This `$\alpha$' lasted for a short time, since observations from 18th June, 2000 to 25th June, 2000
as reported in Naik et al (2002b) always showed a $\rho$ class with gradually increasing burst recurring time-scale.
On the 25th, this trend reversed after this `$\alpha$' class and from the 26th the recurrenced time again
went up.  So the system might have experience perturbations for only a short time.

\begin{figure}
\vbox{
\vskip -3.0cm
\centerline{
\hskip 3.0cm
\psfig{figure=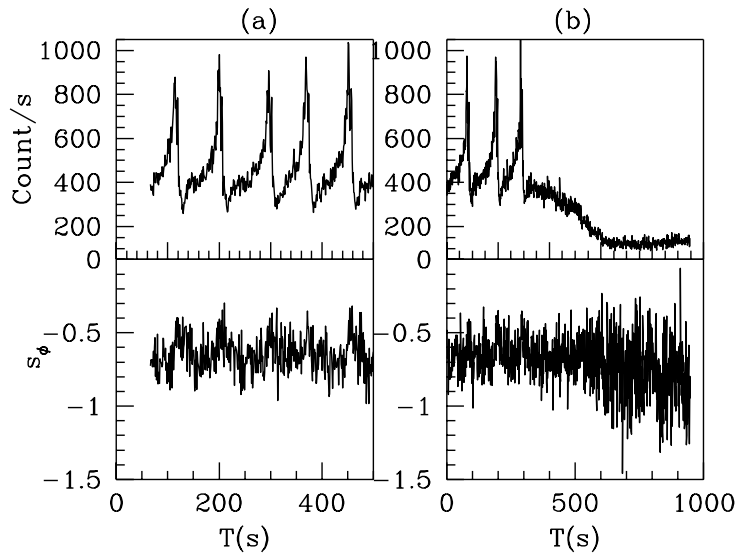,height=16truecm,width=16truecm}}}
\vspace{-2.0cm}
\caption[] {Two successive orbital IXAE data showing a class transition on the 25th of June, 2000. In (a), the class
is similar to $\rho$ but the count rate is higher and the recurring 
time-scale between bursts is large ($\sim 100$s). In (a), the spectral index is less noisy but in the beginning of (b)
it becomes noisy though superficially it is still in `$\rho$' class. After transition it went to the so-called $\alpha$
class and $s_\phi$ becomes very noisy. 
}
\end{figure}

The change of class is also reflected in Fig. 6 where the time dependence 
of the power density spectra (PDS) is plotted. Along the Y-axis, the frequency ($\nu$)
of the PDS is presented. The power (P) itself is marked on the contours 
plotted: solid curve, dotted curve and dashed curves are for $log(P)=-0.5$, $-0.8$ and $-1.2$
respectively. Note that the highest power remains at around $log(\nu)\sim -2$ in the $\rho$
state. A weaker peak occurs at around $log(\nu) = -0.85$. However, after the transition, 
the dominant frequency seems to be at around $log(\nu)\sim 0.4$ which corresponds to 
$\nu \sim  2.5$Hz. 

\begin{figure}
\vbox{
\vskip -4.0cm
\centerline{
\hskip 2.0cm
\psfig{figure=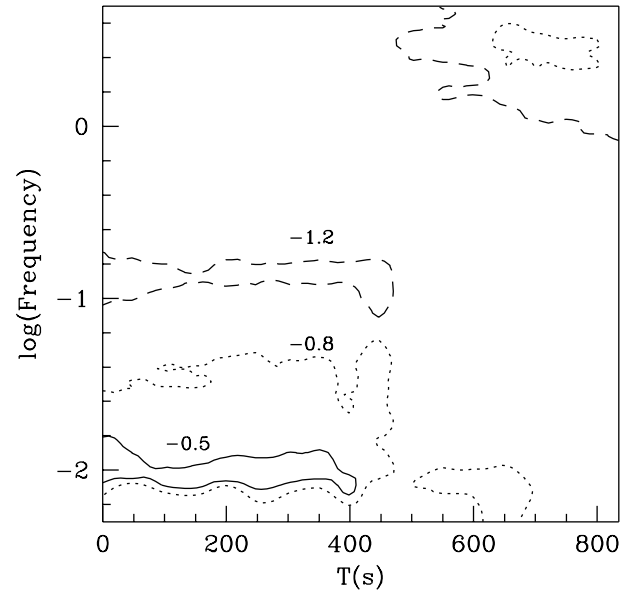,height=15truecm,width=15truecm}}}
\vspace{-1.0cm}
\caption[] {Time dependence of the power density spectrum for
the  medium mode ($0.1$s time resolution) IXAE observations presented in Fig. 5b. Along X-axis is the time
after the observation started. The contours of constant power (in logarithmic scale)
are plotted (marked on the contours). A strong peak at $log(\nu_{QPO}) \sim -2$  
in the first half signifies the Low Frequency QPO (LFQPO) in the $\rho$ state and
another weak peak at $\sim 0.1$Hz (break frequency, see text) is also present. 
After the transition, the peak occurs at $\sim 2.5$Hz.}
\end{figure}

\section{Possible nature of the accretion flow emerging from class transitions}

The first and the most important point to note is the variation in the count rate in the pre-transition 
period and the duration of a transition. The variation in the count rate points to the variation in 
the accretion rate while the duration gives an indication of the infall time. Details of the possible 
nature of the flow geometry during transition will be discussed elsewhere (Nandi et al.
in preparation). Given that there is a gap of more or 
less $80$ minutes in between two successive observations of IXAE, the duration $T_d$ could be at the 
most $\sim 3 - 5$ hours i.e, $10,000-20,000$s. This is short even for a free falling gas from 
the outer edge of the disk located at $r_d \sim 1.5 \times 10^6 r_g$, where $r_g=2GM/c^2$ is the 
Schwarzschild radius of the central black hole of mass $M \sim 14 \pm 4 M_\odot$ (Greiner, 
Cuby and McCaughrean, 2001) since this is around 
$$
T_{infall} \sim r_d^{3/2} (r_g/c) \sim (\frac{r_d}{1.5\times 10^6})^3/2 (\frac{M}{14M_\odot}) {\rm s}
\sim 2.6 \times 10^5 {\rm s} .
\eqno{(3)}
$$
The viscous time for a Keplerian disk of similar size must be at least ten to a hundred times larger,
i.e, few $\times 10^6$s for any reasonable viscosity. This indicates that if the transition takes place
in $\sim 10^4$s, the accretion flow must be nearly
freely falling, i.e., sub-Keplerian, and must originate from intermediate distances,
rather than from the outer edge, i.e., out of a Keplerian disk through energy deposition
or otherwise. This flow is neither a static corona, nor a flow
which is radiatively less efficient. Smith et al. (2001)
and Smith, Heindl and Swank (2002) indeed found observational signatures of the nearly 
free-falling matter in several black hole candidates which causes dynamical spectral 
state changes. We  thus believe that the variation of the rate of the sub-Keplerian matter 
may be responsible for the class transitions we presented here.

\section{Concluding remarks}

In this paper, we presented several examples of variability class transitions in GRS 1915+105 as observed 
by the Indian X-Ray Astronomy Experiment (IXAE) aboard the Indian Satellite IRS-P3. We also
presented one example from RXTE. We showed that while the signature of a class transition 
in the light curve may be abrupt, the process itself is gradual over a period of about $3-5$ hours
during which the light curve passes through unknown classes. During the transition, the
photon count rates change significantly which indicates changes in the accretion rates. 
In a model-independent way, we argue that probably only the rate of the sub-Keplerian flow changes since 
the duration of transition is $\sim 2 \times 10^4$s, much shorter than the viscous 
time by factor of ten to hundred. During the 
transitions, the photon count rates were found to be abnormal and were rapidly changing.
For instance in Fig. 3a, the X-ray count rate was seen to vary by more than $25$\% in a matter of a few 
minutes in the unknown class. These are indications that nearly freely falling 
(i.e., a low angular momentum) sub-Keplerian flow may  present in the accretion flow of GRS 1915+105,
supporting earlier conclusions of Smith et al. (2001) and Smith, Heindl and Swank (2002) in the 
context of  several other black hole candidates.

This work is supported in part by  Grant No.  SP/S2/K-15/2001 of Department of Science and Technology, Govt. of India. 
The authors thank Prof. P.C. Agrawal for allowing the IXAE data to be placed in the ISRO-sponsored Databank at Centre for 
Space Physics which were analysed in this paper.

{}

\end{document}